\documentclass{wlscirep}

\title{Engineering Calcium Signaling of Astrocytes for Neural-Molecular Computing Logic Gates}

\author[1,2,*]{Michael Taynnan Barros}
\author[2]{Phuong Doan}
\author[2]{Meenakshisundaram Kandhavelu}
\author[3]{Brendan Jennings}
\author[3]{Sasitharan Balasubramaniam}
\affil[1]{School of Computer Science and Electronic Engineering, University of Essex, Colchester, UK.}
\affil[2]{BioMediTech, Faculty of Medicine and Health Technology, Tampere University, P.O.Box 553, 33101 Tampere, Finland.}
\affil[3]{Telecommunication Software \& Systems Group (TSSG), Waterford Institute of Technology (WIT), Ireland}

\affil[*]{m.barros@essex.ac.uk}

\begin{abstract}
The aim of molecular computing is to develop analogue and digital logic systems that are established from chemical interactions of molecules. Prokaryotic cells, and in particular bacteria, have received tremendous attention for molecular computing, mainly due to the ease of programming new functions through genetic circuits that are embedded into their DNA plasmids. This paper proposes the use of Eukaryotic cells, namely astrocytes, to develop logic gates. The logic gates are achieved by manipulating the threshold of Ca$^{2+}$ ion flows between the cells, based on the input signals. Through wet-lab experiments that engineer the astrocytes cells with pcDNA3.1-hGPR17 genes, we show that both AND and OR gates can be implemented by controlling Ca$^{2+}$ signals that flow through the population. A reinforced learning platform is also presented in the paper to optimize two main parameters, which are the Ca$^{2+}$ activation threshold and time slot of input signals $T_b$ into the gate. This design platform caters for any size and connectivity of the cell population, by taking into consideration the delay and noise produced from the signalling between the cells, in order to fine-tune the activation threshold and input signal time slot parameters. To validate the effectiveness of the reinforced learning platform, a Ca$^{2+}$ Signaling-based Molecular Communications Simulator was used to simulate the signalling between the astrocyte cells. The results from the simulation showed that an optimum value for both the Ca$^{2+}$ activation threshold and time slot of input signals $T_b$ is required to achieve optimal computation accuracy, where up to 90\% accuracy for both the AND and OR gates can be achieved with the right combination of values. The reinforced learning platform for the engineered astrocytes to create digital logic gates can be used for future Neural-Molecular Computing chip, which can revolutionize brain implants that are constructed from engineered biological cells. 

\end{abstract}

\begin{document}

\flushbottom
\maketitle

\thispagestyle{empty}

\section*{Introduction}

Synthetic biology has facilitated capabilities to engineer biological cells that can lead to novel applications in medicine as well as in biotechnology \cite{hansen2016synthetic,lapique2017genetic,weinberg2017large}. This engineering process is realized through the synthesis of genetic circuits that results in new cell functions, and an example is controlling cellular intra and inter-communications. Numerous digital-like devices have emerged from synthetic biology, and this includes toggle switches \cite{gardner2000construction}, oscillators \cite{elowitz2000synthetic}, as well as Boolean logic gates \cite{goni2019high}. Boolean logic gates, in particular, has received considerable attention due to their ability to be assembled into logic circuits that can perform computation, which can be used to reconfigure cellular operations for therapeutic purposes, or diagnostic sensing of multiple enzymes that are indicators of diseases \cite{Siuti:2013:nature,Lienert:2014:naturereviewmol} \cite{Bacchus2013}. Examples of boolean logic gates that have been engineered from cells includes AND \cite{Anderson2007}, OR \cite{Stetter:2006:bionetics}, NOR \cite{Tamsir2011}, and XOR \cite{gong2019programmable} gates. 
An essential function in synthetic logic circuits is communication, and this can be short-range between different genes within a circuit, to communication between populations of cells that represent individual gates. 
Therefore, engineering molecular communication between the cells through engineered genetic circuits can not only produce logic gates with multiple computational functions but can enable reconfigurability of the logic operations \cite{Akyildiz:2015:ieeecommag,abbasi2018controlled,kuscu2015fluorescent,martins2019quality}. Molecular communication is an emerging paradigm that aims to characterize as well as engineer biological communication systems using communication engineering theory in conjunction with synthetic biology \cite{Akyildiz:2008:ComputerNetworks,akyildiz2019moving,Galluccio201353,barros2018feed}, and one example is calcium-based molecular communication system. Synthetic circuits to control molecular communications use ligand-responsive transgene systems that can respond to a particular stimulus \cite{ma2016foundations} \cite{Hiratsuka:1999:ieeetrans,Stetter:2006:bionetics}, and this can enable reconfiguration when specific signalling molecules activate the circuit. 

\begin{figure}
        \centering
        \includegraphics[width=0.9\textwidth]{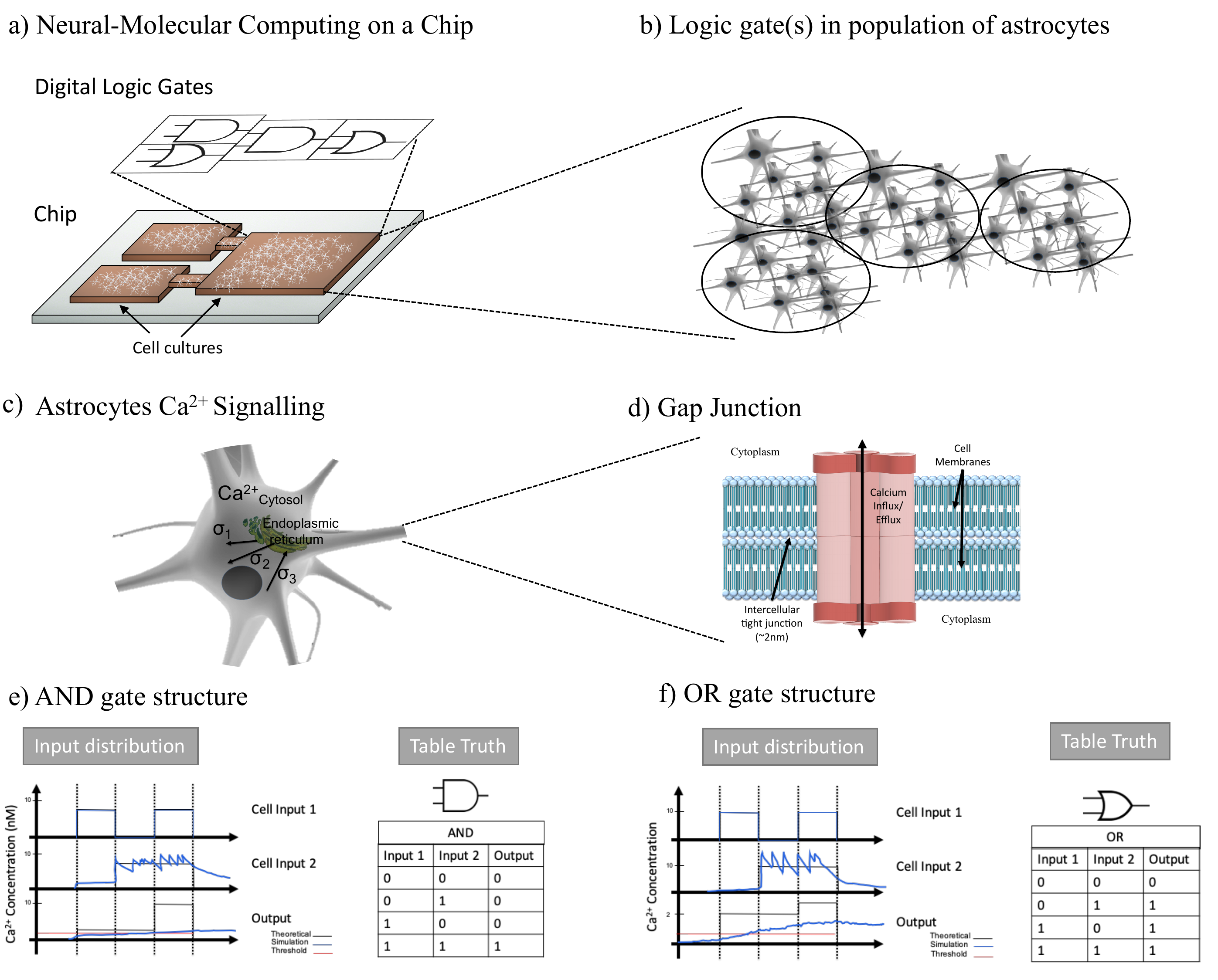}
        \caption{Neural-Molecular computing on a Chip (a), is composed of digital logic gates that are implemented from engineered astrocytes cells (b). The input 1 and 2 astrocyte cells are the incoming Ca$^{2+}$ signals into the gate, which is computed to produce an output signal. The digital gate behaviours are achieved by engineered the threshold of the intracellular Ca$^{2+}$ signalling process (c) and the gap junctions that facilitate cell-cell Ca$^{2+}$ diffusion (d). Simulation illustration of Ca$^{2+}$ signals through cells that represent and AND gate (e) as well as OR gate (f).}
        \label{fig:cal_sig}
\end{figure}


In this paper we attempt to create Boolean logic gates from eukaryotic cells, namely astrocytes (Fig. \ref{fig:cal_sig} (a-c)). 
Astrocytes are glial cells that are found in the brain tissue and have a particular function in supporting stability for the neurons. This stability comes in the form of structural support for neurons, providing nutrients and oxygen supply, pathogen destruction, and removal of dead cells. 
To create logic gates from astrocytes (Fig. \ref{fig:cal_sig} (d)), our approach is based on engineering the Ca$^{2+}$ signalling (Fig. \ref{fig:cal_sig} (a) (b)) between the cells, and this is achieved by using synthetic genes that will set the threshold value for the sensitivity of the intracellular Ca$^{2+}$ signalling. Ca$^{2+}$ signalling is a short-range inter-cellular communication process that uses ions as the signalling molecules between the gap junction connection of the cells. In \cite{Barros2017,Barros:2015:IEETCOM}, an engineered Ca$^{2+}$ signalling-based molecular communication system was analyzed to understand the short-range communication properties, and in particular, the behaviour of ion propagation throughput within a tissue. 
{\color{black}The logic gate structure, depicted in Fig. \ref{fig:cal_sig} (f), includes three populations of cells that each corresponds to a link with a logic gate (2 input links and 1 output link). The two input links of Ca$^{2+}$ signalling populations represents the input into the digital logic gates, which are then transmitted to the output link population with the engineered threshold to perform the logic gate operation. The threshold control of the output link population will determine if it is either an AND or OR gate.} 
The logic circuits built from astrocytes will lead to a new breed of Neural-Molecular Computing on a Chip, as illustrated in Fig. \ref{fig:cal_sig} (e), and will transform implantable brain chips which to date have predominately been developed from non-biological materials.
However, to fully realize the logic circuits built from astrocytes, and rely on  Ca$^{2+}$ signalling as part of its operation, there are many challenges due to the stochastic behaviour of the signalling nature of the cell, and this includes: 

\begin{enumerate}
    \item \textit{Impact from self-regulating spontaneous signalling of Ca$^{2+}$ ions:} The stochastic nature of Ca$^{2+}$ signalling leads to unpredictable stimulation and propagation of Ca$^{2+}$ ions that result in noise that can affect the reliability of the logic operation.
    \item \textit{Inter-cellular Ca$^{2+}$ propagation delay:} Inter-cellular signalling is prolonged compared to conventional CMOS bus lines found in digital logic gates. This property can lead to issues in synchronizing the communication between the cells during the logic operation. Therefore, modelling and characterizing the astrocyte network is critical to ensure the optimal population of astrocytes that will minimize false positive or negative results from the logic gate operation.
    \item \textit{Impact from uncertainty in the astrocyte cells network structure:} 
    {\color{black} The unknown network topology of the astrocyte cells can lead to different delays that can impact the logic operation reliability.}
\end{enumerate}

To address the challenges described above, we developed a reinforced-learning platform that will be used to assist the design of the logic gate from the astrocyte cells. 
The reinforced-learning algorithm analyses the molecular communication of Ca$^{2+}$ ions within the astrocyte network, and through a state value function learning process, adapt parameters that determine the threshold level needed to be engineered in the cells to encode active states of Ca$^{2+}$ ions in the output link. Fig. \ref{fig:cal_sig} (g,h) depict the gate function for a small population of astrocytes, where we show through theoretical simulations the processing of Ca$^{2+}$ signals in the output link population, and compared it to the truth table of the respective logic gate.
In the following, we present a summary of the paper's contribution:

\begin{itemize}
    \item \textbf{Eukaryote cell-based synthetic logic gate:}  In contrast to previous approaches that use prokaryotic cells for developing the gates, we propose a model for developing AND and OR logic gates from astrocyte cells. 
    \item \textbf{In-vitro experimentation:} {\color{black}Experiments performed using hGPR-17 synthetic gene expression in astrocytes as well as two sets of chemical compounds ($MDL29,951$ and $T0510.3657$) that are used to simulate incoming Ca$^{2+}$ signal inputs. The experiment conducted in a petri dish  demonstrates an AND and OR gate operations based on threshold control of the Ca$^{2+}$ signalling.}
    \item \textbf{Reinforced learning platform for logic gates design:} The reinforced learning algorithm uses a closed-loop feedback system to fine-tune the astrocyte cell-cell communication parameters including the activation Ca$^{2+}$ threshold and communication period, that will ensure the engineered population of logic gate cells can be integrated into tissue and operate reliably. This will enable the platform to be used for future practical applications that require engineered astrocytes to perform logic operations. The adaptive process tunes the Ca$^{2+}$ signalling activation threshold to determine the optimal control of ions flow that will result in reliable gate operation. 
    \item \textbf{Accuracy and delay analysis:} Due to the fluctuation behaviour of cell-cell communication with inter-cellular Ca$^{2+}$ signalling, we theoretically analyzed the accuracy of the logic operation, as well as 
 {\color{black}the delay of input flow to the logic gate} using the \textit{static timing analysis} from conventional digital logic circuit theory. 
\end{itemize}

\section*{Methodology}

Our methodology involves both wet-lab experiments as well as theoretical simulations of the astrocyte-based logic gates. 
{\color{black} Engineering logic operations in a network of astrocyte cells must consider the cell-cell Ca$^{2+}$-signalling and its impact on the reliability of the computing functioning. This includes considering their internal signalling pathways and its relation to the engineered threshold, gap junctions probabilities for Ca$^{2+}$ ion propagation, as well as the delay of signals that is dependent on the network connectivity.
The random connectivity of the astrocyte cells network will lead to varying 
 noise, delay and signal fading, which impacts on the reliability performance of the logic gates. 
In order to ensure a high reliability of the logic operation, we use a Reinforced Learning platform illustrated in Fig. \ref{fig:methodology_full}, which will take in culture of astrocyte and based on end-to-end Ca$^{2+}$ signaling through the culture, will fine tune two parameters which are the the optimal Ca$^{2+}$ activation threshold to be engineered into the cells and the optimal transmission period ($T_b$). In this paper, we used a Ca$^{2+}$ Signaling-based Molecular Communications simulator (bottom layer of Fig. \ref{fig:methodology_full}) to simulate the astrocyte culture. The reinforced learning platform utilizes functions, which are designed to adapt system state variables of the inputs by recursively averaging the configuration parameters, and this will result in the optimal Ca$^{2+}$ activation threshold and the transmission period ($T_b$).

This section will present the Reinforced Learning platform that includes the the Ca$^{2+}$ Signaling-based Molecular Communications simulator, as well as the methodology for the wet lab experiments.}

\subsection*{Synthetic Logic Gate Design Platform using Reinforced Learning}
 
\begin{figure}
        \centering
        \includegraphics[width=0.6\textwidth]{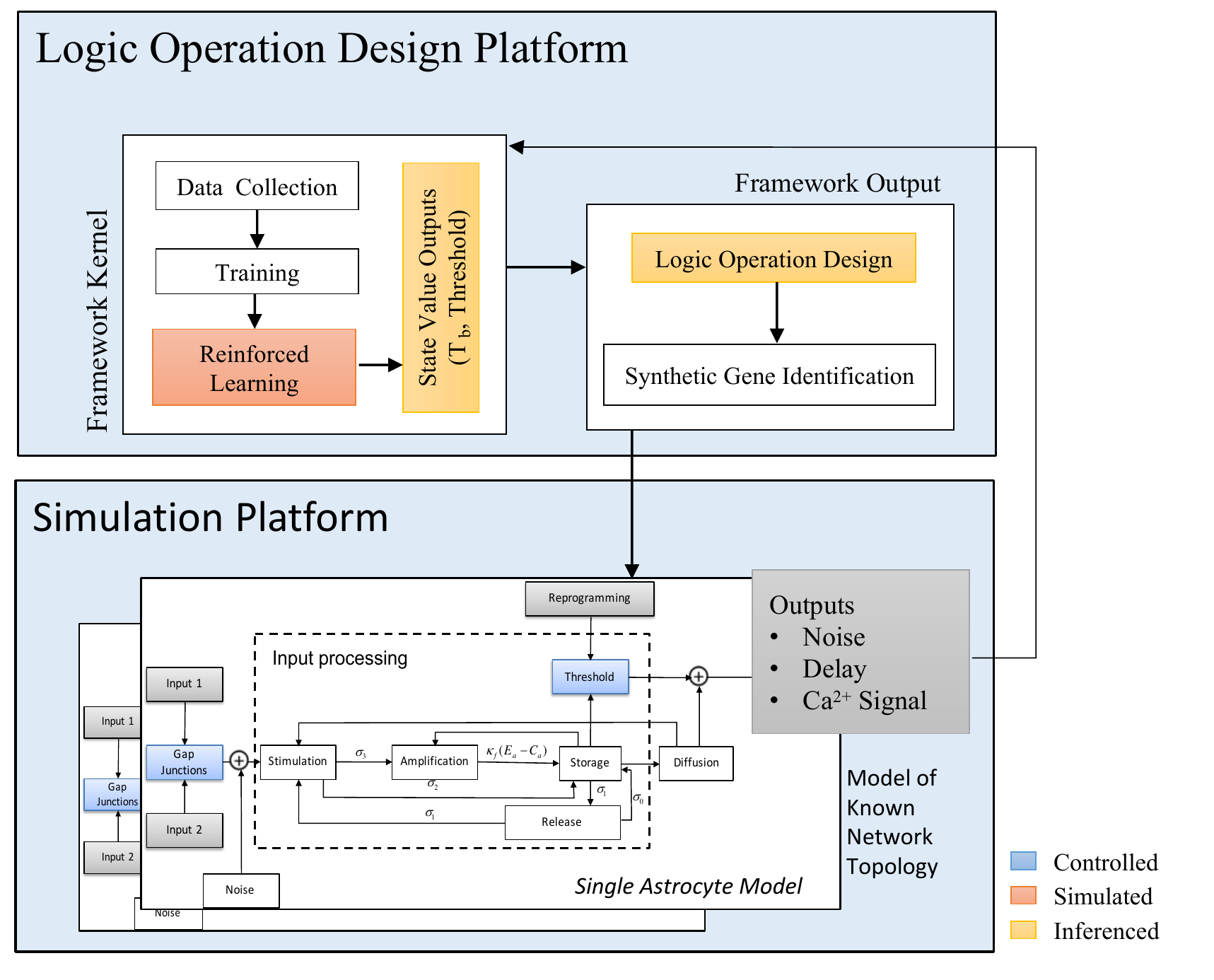}
        \caption{Astrocyte-based logic gate design platform using reinforced learning. The platform is a feedback system for reinforced learning using state value functions that fine tunes the Ca$^{2+}$ activation threshold as well as time slot for input signals $T_b$. The Kernel of the platform considers information such as the delay of signals between cells in the population, noise, transmission period, required logic operation, input signal flow concentration of Ca$^{2+}$ ions, which are fed into the data collection that is used for training. In the output of the platform, synthetic gene transcription is identified for the logic operation based on the defined values of the cellular signalling threshold the is tuned by the kernel state value functions. To validate the platform, a Ca$^{2+}$ signalling-based molecular communications simulator is integrated. The simulator includes models for individual cell's intra as well as intercellular signalling for a defined toplogy. The output population will produce Ca$^{2+}$ signals based on the logic computation. In blue are the controlling blocks that define the threshold for the logic gate operation. The blocks within the "Input Processing" are stages of the Ca$^{2+}$ signalling process in each astrocyte cell.}
        \label{fig:methodology_full}
\end{figure}

{ \color{black} \emph{Reinforced Learning Platform}: The core parts of our proposed platform illustrated in Fig. \ref{fig:methodology_full} are the Kernel and the Output. While the Kernel is responsible for setting the learning rules from the state value functions, the output implements the learning rules that set the Ca$^{2+}$ signalling activation threshold values for the AND or the OR gates. The framework converges the Ca$^{2+}$ signalling activation threshold as well as $T_b$ values based on input data of the astrocyte network to the Kernel (while this will come from the culture directly, in our case we are using the simulator to produce the input data). The Kernel is the most complex part, whereby the processed cellular population input data undergoes the training and the reinforced learning process. 
For simplicity, we used the cellular population input values of the framework as training features directly. As illustrated in Fig. \ref{fig:methodology_full} the simulator of the Ca$^{2+}$ signalling molecular communications in the astrocytes population will refine and converge the values of the optimal Ca$^{2+}$ activation threshold and $T_b$, by minimizing the noise and delay as signals are transmitted through the population, which makes the design agnostic to any network topology of astrocytes. Once the optimal Ca$^{2+}$ activation threshold and $T_b$ values are identified for the specific astrocyte population, a synthetic circuit is designed to stimulate Ca$^{2+}$ signal in the output link once the flow of ions from the two input links reaches the threshold.}

\emph{Ca$^{2+}$ Signalling-based Molecular Communications Model}: The astrocyte cell communication is characterized by both the intracellular as well as the intercellular signalling processes, and the simulator for this signaling process shown in the bottom layer of Fig. \ref{fig:methodology_full}) is described as follows: 
The intracellular Ca$^{2+}$ signalling (Fig. \ref{fig:cal_sig} (b)) is based on the classical Goldbeter et al model \cite{Goldbeter:1990:PNAS}. The model is based on stimulating the IP$_3$ protein, resulting in Ca$^{2+}$ signals generation and release (\textit{Stimulation}). The released IP$_3$ indirectly controls the influx of Ca$^{2+}$ ions to the endoplasmic reticulum and its storage in the cytosolic area (\textit{Storage}). Besides the stimulation process, certain cellular components, such as the endoplasmic reticulum and mitochondria, are also capable of self-generating Ca$^{2+}$ ions (\textit{Amplification}). Finally, the exchange of Ca$^{2+}$ ions is conducted in two ways: cell-cell communication (\textit{Diffusion}) and aleatory exchange of Ca$^{2+}$ to the extracellular space (\textit{Release}). In inter-cellular Ca$^{2+}$ signalling, ions are propagated through the cellular tissues via a physical gate that connects the cytosolic areas of two neighbouring cells, and these gates are called \emph{Gap Junctions}. {\color{black} Fig.\ref{fig:cal_sig} (a) shows how the gap junctions connect two cytosols.} The gap junctions are composed of two \emph{connexons}, one in each connecting cell, which is formed by six proteins called \emph{connexins}. Inter-cellular diffusion only occurs when both connexons open at the same time. 
The voltage-sensitive gap junctions are assumed to have two states of conductance for each connexin: an open state with high conductance and a closed state with low conductance as illustrated in Fig.\ref{fig:cal_sig} (c, d). For astrocytes, the gap junctions are closed around 18\% of the time on average, which probabilistically dictates the cell-cell propagation of Ca$^{2+}$ ions that can subsequently interfere with the logic gate operation \cite{Barros:2015:IEETCOM}. The theoretical modelling of the inter/intra Ca$^{2+}$ signalling with the gap junction diffusion of astrocytes population is presented in the Supplementary Material.

\subsection*{Wet-lab experimental set-up}

{\color{black}The in-vitro experiments aims to determine the sensitivity of the astrocyte cell culture to the induced Ca$^{2+}$ signals. This will determine the validity of the engineered threshold that differentiates between the AND and OR gates. Specifically, we targeted a population of astrocyte cells that have their thresholds  controlled by the hGPR-17 gene and Ca$^{2+}$ signals induced by either the MDL29,951 or T0510.3657 chemical compounds.} The two gates were programmed to induce fluorescent light with increasing Ca$^{2+}$ concentration values as the concentration of ions crosses the threshold to indicate a successful logic operation. The Ca$^{2+}$ signal output of one astrocyte population is the input to a neighbouring population, and this intercellular signalling will be defined by threshold value from the reinforced learning platform. This intercellular signalling activation process can be further explored in future works by having a unique threshold setup for different cell types. 
Our experimental design is based on the approach in \cite{saravanan2018identification}.

\textbf{Cell Culture and hGPR-17 gene expression.} Human astrocytoma cells, 1321N1, were cultured in Dulbecco's modified Eagle's medium with L-glutamine (DMEM-high glucose) (Sigma-Aldrich) supplemented with 10\% (v/v) fetal bovine serum (FBS) (Sigma-Aldrich), penicillin and streptomycin (100U/ml) (Sigma-Aldrich), sodium pyruvate 1mM (Sigma-Aldrich), and amphotericin B 250$\mu g/ml$ (Sigma-Aldrich) and grown at 37 $^{\circ}$C in CO$_2$ incubator. Cells were seeded in a 25 $cm^{2}$, T-25 flask (Fennokauppa) and after 24 hours of incubation, hGPR17 plasmid was transfected with Ca$^{2+}$ phosphate transfection kit (Sigma-Aldrich). The hGPR17 gene was cloned into pcDNA3.1 plasmid, which is a mammalian expression vector \cite{hennen2013decoding}. 3$\mu g$ of pcDNA3.1-hGPR17 plasmid was used to transfect 1321N1.
Post transfection of the plasmid, the media was removed, and fresh DMEM containing 10\% FBS was replaced. 
For the Ca$^{2+}$ time-lapse analysis MMK1 cells, GBM cells derived from patient samples which overexpressed GPR17 were plated in 96-well plates at an initial density of 1$\times$104 cells per well. The cells were incubated overnight to reach around 70\% of confluence. To measure Ca$^{2+}$ level change over time, the cells were incubated with 5 $\mu$M Fura-2 AM (Sigma-Aldrich, St. Louis, MO, USA) for 30 min at 37 $^{\circ}$C. The cells were washed with PBS twice before adding 50 $\mu$L complete medium. Then, 50 $\mu$l of PBS containing 25 $\mu$M concentration of the MDL29,951 or T0510.3657 were added to the wells, and Ca$^{2+}$ changes were measured immediately for 1.5 hours.



\textbf{Quantification of cellular Ca$^{2+}$ signals.} The level of cellular Ca$^{2+}$ was quantified using the Fura2-AM Assay kit (Sigma-Aldrich). 24 hours of post-transfection, transient cell line was collected on 96-well plates at a concentration of 1 $\times 10^5$ cells/well. Cells were incubated with increasing concentrations of signalling molecules, MDL 29951 (Abcam), and T0510.3657 (AKos Consulting \& Solutions Deutschland GmbH) at 37ºC for 2 hours. 10$\mu M$ Fura 2-AM was added to the cells and then assayed for Ca$^{2+}$ accumulation after the 30 minutes of incubation at 37ºC, following the manufacturer's instructions. The difference between the fluorescence level of the control and signalling molecule treated samples were measured using the plate reader (Ascent). In the experimental set up following conditions were used: 1) Cells without the transfection of plasmid, 2) Cells with the transfection of plasmid and without the compound incubation, 3) Cells without the transfection of plasmid and with the incubation of 50 $\mu M$ concentration of compounds, 4.a) Cells with the transfection of plasmid and with the incubation of 25 $\mu M$ of compound and 4.b) 50$\mu M$ of the concentration of compounds. To quantify the changes in Ca$^{2+}$ level, the kit protocol, as given by the vendor, was used. Technical and biological repeats were used to measure fluorescence and were averaged.  The fluorescent signal was measured using a microplate reader (Spark®, Tecan) at two dual excitation/emission wavelengths of 340/510 nm and 380/310 nm.

\section*{Results}

\subsection*{Wet Lab Experiments}
\begin{figure*}[tb!]
\centering
\includegraphics[width=1\textwidth]{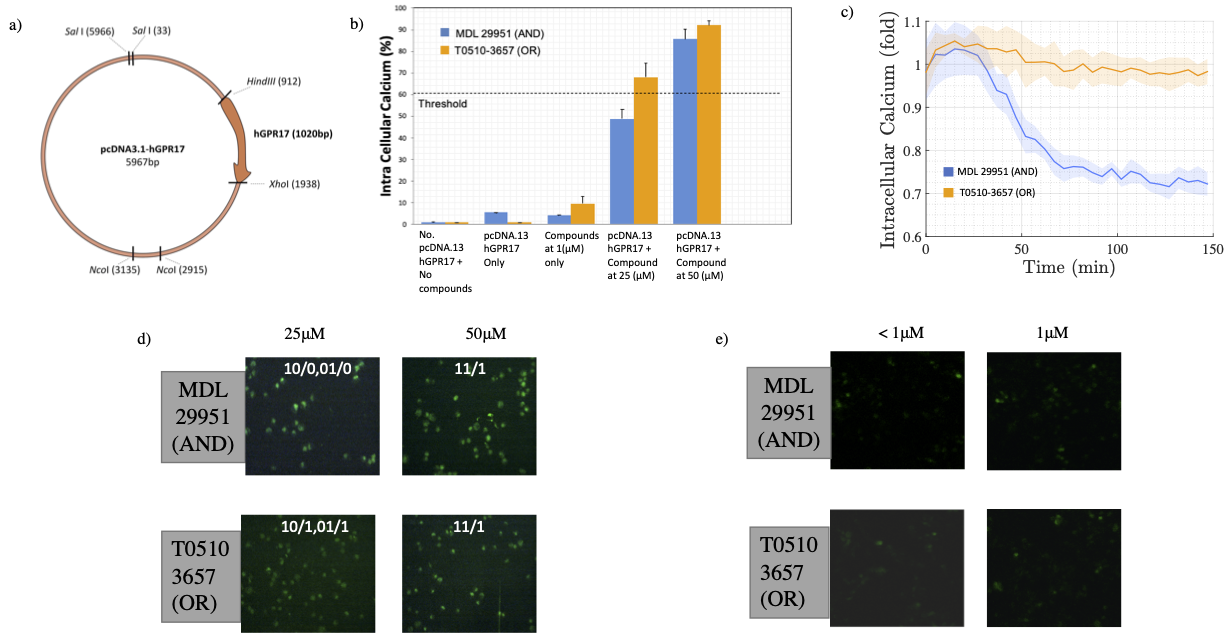}
\caption{Wet lab experiments of the logic gate input control for an astrocyte population in-vitro. (a) The pcDNA3.1-hGPR17 plasmid used to engineer the threshold of the astrocytes in the experiments.  
In (b) the percentage of intracellular Ca$^{2+}$ concentration over five in-vitro setups: no pcDNA3.1-hGPR17 (no gene), pcDNA3.1-hGPR17 (genes only), no pcDNA3.1-hGPR17 with compounds (compounds only), pcDNA3.1-hGPR17 with compounds at 25$\mu M$ (genes and compounds) and pcDNA3.1-hGPR17 with compounds at 50$\mu M$ (genes and compounds). 
Based on these results, we observe the levels of fluctuation of the Ca$^{2+}$ signals that affect the performance of the logic operations. In (c) we show the intracellular Ca$^{2+}$ variation over time for both compounds. 
In (d) we show the increasing Ca$^{2+}$ concentration observed by the fluorescent light effect of in-vitro astrocyte cultures with compound concentrations of 25$\mu M$ and 50$\mu M$ for the AND gate with the $MDL 29,951$ compound and the OR gate with the $T0510.3657$.
{\color{black} Finally, in (e), we show the response of the logic gates with random signals as inputs, to demonstrate that low Ca$^{2+}$ signals will not produce sufficient signals to activate the gate response.}
}
\label{fig:wetlab}
\end{figure*}

{\color{black} Fig. \ref{fig:wetlab} presents the wet lab experiments to demonstrate the AND and OR gates that are engineered from the astrocyte cells. Fig. \ref{fig:wetlab} (a) illustrates the engineered plasmid with the gene pcDNA3.1-hGPR17 insertions that are used to amplify the Ca$^{2+}$ signals for the two gates. The OR gate is a combination of $T0510.3657$ compound added to the gene pcDNA3.1-hGPR17, while the AND gate is a combination of $MDL 29,951$ with the gene pcDNA3.1-hGPR17. The compounds are used to simulate the incoming Ca$^{2+}$ signals from input 1 and 2, where 25$\mu$M will represent only a single 1 input, and the 50$\mu$M will represent two 1 inputs. In order to evaluate the effectiveness of the logic operation, Fig. \ref{fig:wetlab} (b) illustrates the different quantity of Ca$^{2+}$ produced with respect to the different amounts of compounds added. The cases of non-engineered cells with no compounds (no incoming Ca$^{2+}$ signal inputs), only engineered cells, and non-engineered cells with only compounds produces a small quantity of Ca$^{2+}$ signals. However, the combinations of engineered cells with compounds (both 25 and 50 $\mu$M) presents high levels of Ca$^{2+}$ signals for both the AND and OR gates. The threshold determines if a 1 output will be produced depending on the type of gates as well as input. In the case of 25$\mu$M, which is a single 1 input, the AND gate will produce Ca$^{2+}$ signals that don't reach the threshold, while in the case of the OR gate the output will surpass the threshold. However, in the case of 50$\mu$M, where there are two input 1, we can observe that the quantity of Ca$^{2+}$ ions produced are above the threshold. Fig. \ref{fig:wetlab} (c) presents the changes in overall stability and fluctuations of the Ca$^{2+}$ signals with respect to time and shows that the AND gate will have a more stable production of the signals compared to an OR gate that will result in signals that fade after a certain period. The fluctuation of Ca$^{2+}$ signals is found throughout all configurations and logic operations, producing an average variation of 2.7\% of internal signalling capacity, with a peak at pcDNA3.1-hGPR17 with compounds of 25$\mu M$ with 6\% fluctuation for both compounds. The regulatory Ca$^{2+}$ intracellular mechanisms associated with the cell-cell communication can produce random fluctuations. Fig. \ref{fig:wetlab} (d) presents the fluorescent output of the cells based on the input signals (xx/y refers to the x being the input and y being the output). The result shows that the 25 $\mu$M produces a certain level of output, but this is lower than in the case of 50 $\mu$M when the two input 1 are applied to the gate. }{\color{black} Finally, Fig. \ref{fig:wetlab} (e) presents the response of the logic gates with random signals as inputs with concentrations below 1$\mu$M and with 1$\mu$M. The random signals appear to cause no effects in the logic gates operations.}



\subsection*{Simulation Experiments}
\begin{figure}
        \centering
        \includegraphics[width=1\textwidth]{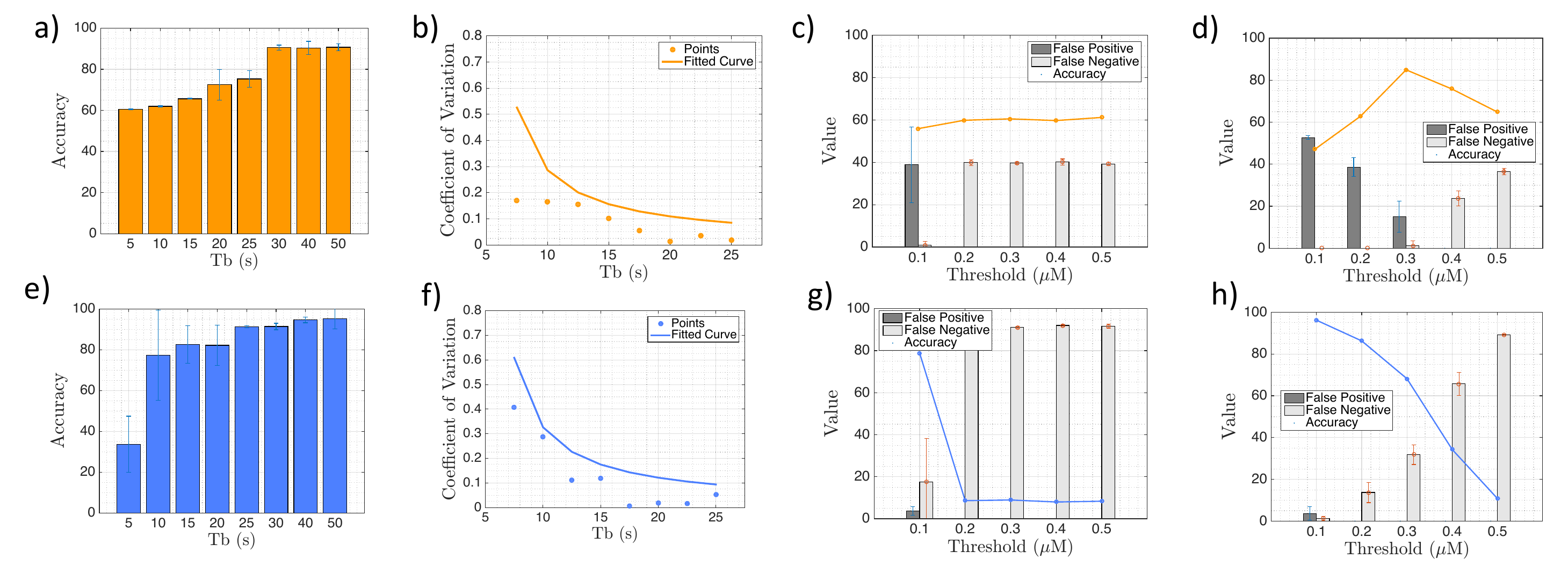}
        \caption{Accuracy analysis of a small astrocyte population with AND (top row) and OR (bottom rows) logic gate operation. The accuracy of the AND (a) and OR (e) logic gate operation over the pulse period $T_b$. Fluctuation levels of Ca$^{2+}$ versus the $T_b$ variation and the obtained regression model curve for AND (b) and OR (f) gates. Accuracy, false positive and false negative results with respect to varying the threshold values for low values of $T_b$ for AND (c) and OR (g) gates. Accuracy, false positive and false negative results with respect to  varying the threshold values for high values of $T_b$ for AND (d) and OR (h) gates.}
        \label{fig:performance}
\end{figure}
 
Fig. \ref{fig:performance} presents results on the logic computing simulation accuracy for the astrocyte cells AND as well as OR gates. The accuracy analysis of a small population of astrocytes AND logic gate is shown in the top row of Fig \ref{fig:performance}. The aim of our analyses is to understand the impact that variation of {\color{black}  Ca$^{2+}$ activation threshold and the $T_b$ of input signals}
will have on the gate's computing reliability under low noise effects.
Fig \ref{fig:performance}(a) shows 
directly increasing $T_b$  benefits the accuracy,  giving peak performance around $90 \%$ for the AND logic operation. This is due to the impact of longer duration of Ca$^{2+}$ signals that can exist in the cell population, as shown in Fig. \ref{fig:wetlab} (c), especially for AND gate, resulting in improved results when higher values of $T_b$ are used. Ca$^{2+}$ signalling fluctuations are represented as the statistical errors of the temporal series shown in Fig \ref{fig:performance} (b), which a simple regression curve shows have an inverse relationship with the $T_b$. Fig \ref{fig:performance} (c) shows the accuracy, false positive and false negative results for variations in the threshold values for a low level of $T_b$. As shown in the results, the accuracy is not affected by the threshold variation for low $T_b$, and stabilizes around $55-60\%$, while false positive has an average of $8\%$ and false negatives has an average of $16\%$. Low values of $T_b$ results in high fluctuation, as can be seen in Fig \ref{fig:performance} (b), which even with different thresholds does not affect the logic operation accuracy. This is contrary to Fig \ref{fig:performance} (c) and (d), which presents the accuracy, false positive and false negative results for variations in the threshold values for high values of $T_b$. An optimal point is observed when the threshold value is around $30\%$ of the intracellular signalling capacity, with accuracy at $80\%$, false-positive at $15\%$ and false-negative at $2\%$. Higher thresholds values are likely due to the interference error in the output due to the increase in false negatives effects that are caused by Ca$^{2+}$ concentration fluctuations. The data from all the results presented in Fig \ref{fig:performance} (a) (b) (c) and (d) that was input into the reinforced learning algorithm, resulted in the optimum value of $T_b = 30s$ and optimal Ca$^{2+}$ activation threshold of $ 0.3 \mu M$, which resulted in an accuracy of 90\% with a minimum percentage of false positives and false negatives.


The accuracy analysis of a small population of astrocyte OR logic operation is presented in the bottom row of Fig. \ref{fig:performance}. 
The results show that higher levels of fluctuations in Fig. \ref{fig:performance} (e) and (f), does not inhibit the OR logic operation from reaching $95\%$ of peak accuracy. However, the high level of fluctuations impacts on the results in Fig. \ref{fig:performance} (g) and (h), where higher false negatives are found (average of $75\%$ for Fig. \ref{fig:performance} (g) and $41.4\%$ for (h)). At the same time, the increase in the threshold values appears to impact on the overall gate performance. Higher levels of fluctuations in low values of $T_b$s are responsible for this effect due to the increase in the false negatives that results from high values of $T_b$. When this data is input into the reinforced learning platform, the optimum value outputs are $T_b = 50s$, and optimal Ca$^{2+}$ activation threshold of $0.1 \mu M$, and this results in an accuracy of $98\%$ with a minimum percentage of false positives and negatives.




\begin{figure}
        \centering
        \includegraphics[width=1\textwidth]{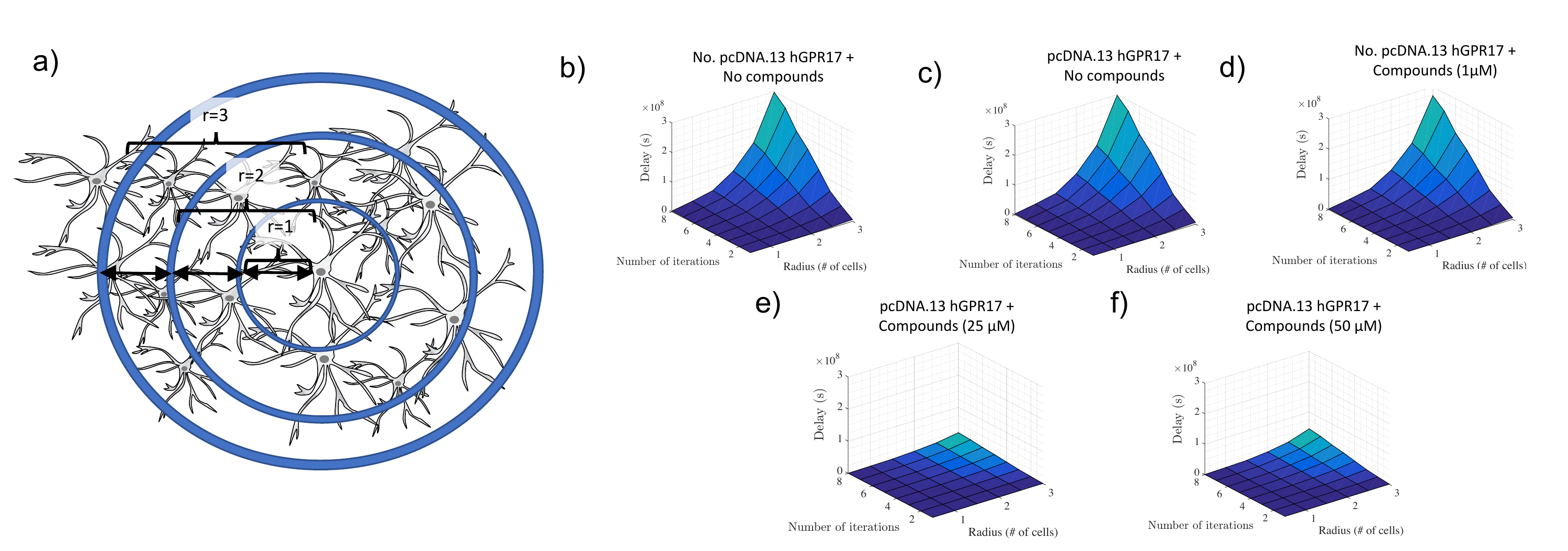}
        \caption{Static timing analysis for delay performance of the input population. (a) The static timing analysis of the input population with different reach radius $r$. (b) Delay in seconds for eight different values of $r$ over 0 to 100 logical operations for no pcDNA3.1-hGPR17. (c) Delay results for pcDNA3.1-hGPR17 only. (d) Delay results for no pcDNA3.1-hGPR17 but with compounds. (e) Delay results for pcDNA3.1-hGPR17 with compounds at 25$\mu M$. (f) Delay results for pcDNA3.1-hGPR17 with compounds at 50$\mu M$.}
        \label{fig:and_delay}
\end{figure}


{\color{black} Our simulations also include the static timing analysis to determine the  Ca$^{2+}$ signal propagation delay through the astrocyte population and its impact on the logic gate operation. The simulation considers the data from the wet lab experiments. Fig. \ref{fig:and_delay} illustrates the topology of astrocytes population that is used for simulating the signal propagation through the output link. As shown in the figure, the topology is based on varying radius $r$ of the astrocyte population for the input link. 
Fig. \ref{fig:and_delay} (b) presents the delay results for the non-engineered cell and shows that as the number of operations increase, this also increases the delay with an increase in population radius $r$. Fig. \ref{fig:and_delay} (c) presents the delay results for the cells that are engineered with the pcDNA3.1-hGPR17 genes with no compounds applied, while Fig.  \ref{fig:and_delay} (d) presents the delay results for non-engineered cells but with the compounds applied. Fig.  \ref{fig:and_delay} (e) presents the delay results for the engineered cells with pcDNA3.1-hGPR17 applied with compounds at 25$\mu M$, and  Fig.  \ref{fig:and_delay} (f) with compounds at 50$\mu M$. 
The high delay as signals propagate through the population in Fig. \ref{fig:and_delay} (b) (c) and (d) is due to the propagation of normal levels of Ca$^{2+}$ ions that are in each cells. However, in the case of Fig. \ref{fig:and_delay} (e) and (f), the Ca$^{2+}$ ions are amplified, and this leads to a larger quantity of concentration that is pushed from cell to cell, resulting in a higher speed of propagation, leading to lower end-to-end delay. Based on fast increase response of many compounds, the results show that the delay has decreased by 90\%, compared to the natural Ca$^{2+}$ signalling.  
While our wet-lab experiments have used the compounds to simulate and induce Ca$^{2+}$ signalling, the delay analysis has shown that the compounds can be used in conjunction with the cell culture to improve the performance of the logic operations for the astrocyte-based logic gates.}

\section*{Discussion}


Our study has found that Ca$^{2+}$ fluctuations are the main source of noise in the astrocyte-based logic gates, as observed in both the wet lab experiments (Fig. \ref{fig:wetlab} b and c) and simulations (Fig. \ref{fig:performance} b). These fluctuations are caused by both the Ca$^{2+}$ intra and intercellular signalling. In the case of intercellular signalling, the noise is dependent on the topology of the astrocyte cell population as Ca$^{2+}$ ions can randomly propagate between the cells in the population. We also know from multiscale analysis  \cite{barros2018multi} that even single-cell irregularities can result in random fluctuations of Ca$^{2+}$ propagation. These noise and random fluctuations can result in unreliable logic operation. Moreover, as shown in both Fig. \ref{fig:wetlab} (b) and Fig. \ref{fig:performance} (a) and (e), the relationship between the threshold as well as the $T_b$ can lead to false results in the logic gate computation. Accuracy can go up to above 90\% levels when $T_b$ is higher than 30s for the AND gate, and for 25s in the OR gate. This is because a fluctuation is shown to decrease when increasing the $T_b$ as shown in Fig. \ref{fig:performance} (b) and (f). However, as shown in Fig. \ref{fig:performance} (c) and (d), the decision of a threshold is dependent on the impact of the output accuracy, false positive and false negatives results, topology structure, the position where the logic gates are placed as well as the system dynamics. The usage of a reinforced-learning approach to decide parameters such as the threshold and $T_b$ can lead the system to optimum results when information about the network topology is not available. This is where the benefits of the reinforced-learning platform comes in, where it determines the optimum $T_b$, by analyzing the propagation noise that is transmitted through the molecular communication simulator until it converges to a value resulting in the least amount of noise, irrespective of the topology. Our experiments in Fig. \ref{fig:wetlab} shows that different compounds amplify the Ca$^{2+}$ signalling for the output of the logic gate, allowing us to set different thresholds for the AND and OR gates. This also means that the topology of the input links plays a role in ensuring that the optimum Ca$^{2+}$ signals should flow into the output population in order to obtain accurate results from the logic computation.
The results from the experiments that used the compounds for simulating the Ca$^{2+}$ signals of the input links was used in the simulations to determine the impact of Ca$^{2+}$ ion concentration propagation on the static delay analysis of the logic operation (Fig. \ref{fig:and_delay}). The simulation has shown that higher concentration of Ca$^{2+}$ propagated between the cells, leads to faster diffusion, which lowers the delay that can lead to higher iterative numbers of logic computations. Therefore, the design for Neural-molecular computing chips can include substrates with the compound mixed with the engineered astrocyte culture, to further amplify the Ca$^{2+}$ ion production as well as propagation. 


Majority of biomolecular computing techniques \cite{Auslander2012,Ye2013,Bacchus2013} developed to date rely on the DNA transcription and translation processes, which limits their operation for future in-vivo applications. This is because it will require insertion of complex genetic circuits into the cells that can result in gene expressions that can be damaging to its biological environment, possibly affecting the tissue homoeostasis  \cite{Pucci:2000:neoplasia}. While there are benefits through the use of cell-free expression techniques, where the machinery are not required to be embedded in a living cell, the operation can be unreliable when all components are required to work together within a liquid environment.  Our approach can partially eliminate these issues with 1) embedding simpler synthetic gene into the cell's genome, and 2) provide a new approach for brain bio-electronics that utilizes engineered astrocytes, which is only based on manipulating the flow of Ca$^{2+}$ ions and thresholds to achieve gate behaviours. The benefit of using astrocytes for the Neural-molecular computing is the ease of integrating it into the brain tissue, where they can easily connect with natural neurons in order to receive incoming signals as well as produce output signals. 
Our work lays the foundation for Neural-molecular computing chips that can embed logic circuits built from gates of engineered astrocytes. Therefore, future work will need to investigate how the astrocyte-based logic gates can be connected into a circuit \cite{Menolascina20122122}. The Neural-molecular computing on a chip that houses the engineered astrocytes can be designed and constructed from biocompatible material (e.g., polymers), avoiding the need for silicon technology to perform the computing.

\section*{Conclusion}

The vision of molecular computing is to perform unconventional computing using biological systems, and in particular through the interaction of molecules produced by cell machinery. Over the years, many molecular computing approaches have been developed, using DNA, where computing functions is achieved through multiple DNA molecules interacting, as well as using cells, such as bacteria. In this paper, we take an alternative approach where molecular computing is achieved through the engineering of Eurokaryotic cells, and in particular, astrocytes. By engineering the threshold of Ca$^{2+}$ ions that flow between the cells, AND and OR gates can be developed. The paper first demonstrated through wet lab experiments AND and OR gates that can be developed using  hGPR-17 synthetic gene expression, with incoming Ca$^{2+}$ signals simulated from chemical compounds ($MDL29,951$ and $T0510.3657$) added to the culture. The results showed that AND and OR gate behaviour can be achieved, provided that the threshold is set accordingly, and this threshold should produce an output Ca$^{2+}$ signal. The paper also presented a reinforced learning platform for logic gate design that is agnostic to any cell culture and can be used to determine the optimum Ca$^{2+}$ activation threshold and input transmission period $T_b$. The validation was performed using a Ca$^{2+}$-signalling based molecular communication simulator. The simulation showed that for any type of input topology of astrocyte network, there is an optimum value for the  Ca$^{2+}$ activation threshold and input transmission period $T_b$, and this was validated through the reinforced learning platform. The future work can use the reinfoced learning platform to design the timing of the input signals as well as the activation threshold for any type of cell culture. The work presented in this paper lays the foundation for future Neural-Molecular Computing on a chip that is constructed from biological cells that perform computing functions, minimizing the need for silicon technology. This, in turn, can result in a game-changer for future brain implants that are controlled and operated from molecular computing logic circuits.

\section*{Acknowledgements}

We are grateful to thank Prof. Dr Evi Kostenis, Institute for Pharmaceutical Biology, the University of Bonn for the kind gift of the pcDNA3.1-hGPR17 plasmid. M. T. Barros is funded by the European Union’s Horizon 2020 research and innovation programme under the Marie Skłodowska-Curie grant agreement No 839553.

\section*{Appendix}

Here we present the mathematical framework used in the paper. The code of the simulations described here can be found in \url{https://github.com/michaelbarros/astrocytes_population_sim}

\subsection*{Computational model of astrocytes intracellular signalling}

In this section we introduce a model describing Ca$^{2+}$
oscillations in astrocytes that was proposed by Lavrentovich and
Hemkin~\cite{Lavrentovich:2008:jtheobio}. The model is in accordance with experimental
observation \cite{Lavrentovich:2008:jtheobio}. We have Ca$^{2+}$ pool storage models, which
includes: Ca$^{2+}$ concentration in the cytosol ($C_a$)
(Eq. \ref{eq:xa}); Ca$^{2+}$ concentration in the endoplasmic
reticulum ($E_a$) (Eq. \ref{eq:ya}); and IP$_3$ concentration
($I_a$) (Eq. \ref{eq:za}). They are represented by the following
equations:
\begin{equation}\label{eq:xa}
\frac{dC_a}{dt} = \sigma_{0} - \kappa_{o}C_a + \sigma_{1} - \sigma_{2} + \kappa_f (E_a-C_a)
\end{equation}
\begin{equation}\label{eq:ya}
\frac{dE_a}{dt} = \sigma_{2} - \sigma_{1} - \kappa_f (E_a-C_a)
\end{equation}
\begin{equation}\label{eq:za}
\frac{dI_a}{dt} = \sigma_{3} -\kappa_{d}I_a
\end{equation}
\noindent where $\sigma_{0}$ is the flow of Ca$^{2+}$ from the extracellular space into the cytosol \footnote{This term can be extended into a voltage dependent term, called voltage-gated Ca$^{2+}$ channels, which was further studied in \cite{zeng:2009:biophysical}.}, $\kappa_{o}C_a$ is the rate of Ca$^{2+}$ efflux from the cytosol to the extracellular space, $\kappa_f (E_a-C_a)$ is the leak flux from the endoplasmic reticulum into the cytosol and $\kappa_{d}I_a$ is the degradation of IP$_3$.

The $\sigma_{1}$ term (Eq. \ref{eq:v3astro}), models the Ca$^{2+}$ flux from the endoplasmic reticulum to the cytosol via IP$_3$ stimulation. In common with excitable and non-excitable cells, this mechanism directly affects the cytosolic concentration of Ca$^{2+}$. It is represented as:
\begin{eqnarray}\label{eq:v3astro}
\sigma_{1} & = & 4\Sigma_{M3}\frac{\kappa^{n}_{C1} C_a^n}{(C_a^n+\kappa^{n}_{C1})(C_a^n+\kappa^{n}_{C2})}\nonumber\\
& & .\frac{I_a^m}{\kappa^m_I + I_a^m}(E_a-C_a)
\end{eqnarray}
\noindent where $\Sigma_{m3}$ is the maximum flux value of Ca$^{2+}$ into the cytosol, $\kappa^{n}_{C1}$ and $\kappa^{n}_{C2}$ are the activating and inhibiting variables for the IP$_3$ 
and the $m$ and $n$ are the Hill coefficients.

The efflux of Ca$^{2+}$ from the sarco(endo)plasmic reticulum to the endoplasmic reticulum is modelled as $\sigma_{2}$:
\begin{equation}\label{eq:vserca}
\sigma_{2} = \Sigma_{M2}\frac{C_a^2}{\kappa^2_2 + C_a^2}
\end{equation}
\noindent where $\Sigma_{M2}$ is the maximum flux of Ca$^{2+}$ in this process.
Finally, $\sigma_{3}$ describes IP$_3$ generation by the Phosphoinositide phospholipase C (PLC) protein:
\begin{equation}\label{eq:vplc}
\sigma_{3} = \Sigma_p\frac{C_a^2}{\kappa^2_p + C_a^2}
\end{equation}
\noindent where $\Sigma_p$ is the maximum flux of Ca$^{2+}$ in this process, and $p$ is the Hill coefficient.

\subsubsection*{Gap Junctions Model} 

A stochastic model of gap junction behaviour was introduced by Baigent et al. \cite{Baigent:1997:jtheobio} and first studied for molecular
communication by Kilinc and Akan \cite{Kilinc:2013:nanotech}. The model
considers voltage-sensitive gap junctions which are assumed to have two states of
conductance for each connexin: an open state with high conductance and a
closed state with low conductance. Based
on this, we consider four basic combinations of states from each connexin
of the connexon:
\begin{itemize}
\item \textit{State HH}: Both gates are in a high conductance state. This probability is denoted by $p_{HH}$;
\item \textit{State HL}: One gate is in a high conductance state, and the other is in a low conductance state. This probability is denoted by $p_{HL}$;
\item \textit{State LH}: One gate is in a low conductance state, and the other is in a high conductance state. This probability is denoted by $p_{LH}$;
\item \textit{State LL}: Both gates are in a low conductance state. This probability is denoted by $p_{LL}$.
\end{itemize}

Experimental validation of the model indicated that the $LL$ state appears
to present very low occurrence rates \cite{Bukaukas:2013:biophycs}, thus
we neglect that state here. Thus, the probabilities should follow:
\begin{equation}
p_{HH} + p_{HL} + p_{LH} = 1
\end{equation}

\noindent Moreover, $p_{HH}$, $p_{HL}$ and $p_{LH}$ are interrelated as follows:
\begin{equation}\label{eq:phl}
\frac{dp_{HL}}{dt} = \beta_1 (\vartheta_j)\times p_{HH} - \alpha_1(\vartheta_j) \times p_{LH}
\end{equation}
\begin{equation}\label{eq:plh}
\frac{dp_{LH}}{dt} =  \beta_2 (\vartheta_j)\times p_{HH} - \alpha_2(\vartheta_j) \times p_{HL}
\end{equation}
\noindent where the control of the gap junctions permeability is mediated through the potential difference of the membrane of two adjacent cells ($\vartheta_j$), the gate opening rate is $\alpha$ and gate closing rate is $\beta$. The terms $\alpha_1(\vartheta_j)$, $\alpha_2(\vartheta_j)$, $\beta_1(\vartheta_j)$ and $\beta_2(\vartheta_j)$ are defined as:
\begin{eqnarray}
\alpha_1(\vartheta_j) = \lambda e^{-A_\alpha (\vartheta_j - \vartheta_0)}\\
\alpha_2(\vartheta_j) = \lambda e^{A_\alpha (\vartheta_j + \vartheta_0)}\\
\beta_1(\vartheta_j) = \lambda e^{A_\beta (\vartheta_j - \vartheta_0)}\\
\beta_2(\vartheta_j) = \lambda e^{-A_\beta (\vartheta_j + \vartheta_0)}
\end{eqnarray}
\noindent where $\vartheta_0$ is the junctional voltage at which the opening and closing rates of the gap junctions have the same common value $\lambda$, and $A_\alpha$ and $A_\beta$ are constants that indicate the sensitivity of a gap junction to the junctional voltage.

\begin{table}[tb]
	\centering
	\caption{Experimental variable values for Cx43 of astrocytes \cite{Baigent:1997:jtheobio}~\cite{Valiunas:2000:circres}~\cite{Moreno:1995:americanphys}~\cite{Valiunas:2002:genphysio}.}
	\label{tb:paramvalues}
	\begin{tabular}{ c c }
	    \toprule
	    Variable & Value  \\
	    \midrule
	    \midrule
	    $\lambda$ & 0.37\\
	    $\vartheta_j$ mV & 90\\
	    $\vartheta_0$ mV & 60\\
	    $A_\alpha$ $\mbox{(mV)}^{-1}$ & 0.008\\
	    $A_\beta$ $\mbox{(mV)}^{-1}$ & 0.67 \\
	    \bottomrule
	  \end{tabular}
\end{table}

\subsection*{Simulation}

We consider a cellular tissue space ($S$) composed of $I \times J
\times K$ cells ($c$), where $c_{i,j,k}$ ($i=1 \dots I$; $j=1, \dots
J$ and $k=1, \dots K$) denotes an arbitrary cell in the tissue. The
cells are connected with a maximum of six neighbouring cells. In the
case of the excitable and non-excitable cells, the organisation
of the cells is assumed to be a layered lattice. However, for
astrocytes, the organisation is going to depend on the type of topology
connection. We use a simple regular connection to perfectly match our
lattice model, that is based on the study of astrocytes topologies
\cite{Lallouette:2014:compneuroscience}.

Consider that each cell contains a set of internal reactions of 
P1 and P2 pools. 
Each reaction and pool for a specific cell type were defined in the previous section. The
stochastic solver computes the values of each pool over time,
selecting and executing scheduled reactions. The pool will be negatively
or positively affected by a constant $\alpha$ when a specific reaction is
executed.


Modelling diffusion in a cellular tissue area captures the temporal-spatial dynamics of intercellular Ca$^{2+}$ signaling. We use Ca$^{2+}$ concentration difference to model this temporal-spatial characteristic, as follows \cite{Nakano:2010:Ieeenanobio}:
\begin{equation}\label{eq:diffusion_gjs}
Z_{\Delta} (i,j,k,n,m,l) = \frac{D}{v}(|Z_{n,m,l} - Z_{i,j,k}|) \times p_{(.)}
\end{equation}
\noindent where $n$ $\in$ $(i-1,i+1)$, $m$ $\in$ $(j-1,j+1)$, $l$ $\in$
$(k-1,k+1)$, $D$ is the diffusion coefficient, $v$ is the volume of the
cell, and $Z_{\Delta}$ is the difference in Ca$^{2+}$ concentration
between the cells. $p_{(.)}$ is the probability of the gap junction
opening and closing. Based on the gap junction probabilities, we define 
three different diffusion reactions for each cellular connection. Such reactions
are the multiplication of the probabilities ($p_{HH}$, $p_{HL}$ and $p_{LH}$) with the
regular cell-to-cell diffusion probability.


\subsubsection*{Stochastic solver}

We present in this section a stochastic solver, which determines the quantity of each pool over time. 
At each time step, the Gillespie algorithm \cite{gillespie:1997:phychems} is executed to select a random cell and a random internal reaction of that cell, also scheduling a time step ($t$) to each one of them.

The process of executing one of the distinct reactions in $R$ requires a scheduling process divided in two phases---selecting a reaction and selecting a time step. Each reaction is allocated a reaction constant ($a_r$). 
Considering that $\alpha_0$ is the summation of all $a_r$ in $R$, the next reaction chosen $r_u$ will be:
\begin{equation}
r_u = \mbox{MAX}\left\{\frac{a_{r_j}}{\alpha_0} = \frac{a_{r_j}}{\sum\limits_{j=1}^{|R|} a_{r_j}}\right\}, u \in \mathbb{N}, u \in R
\end{equation}
\noindent which follows the \textit{roulette wheel selection} process, which selects the events based on their probability values. However, $u$ must satisfy the following restriction:
\begin{equation}
\sum\limits_{j=1}^{u-1}  \frac{\alpha_{r_j}}{\alpha_0}< \rho_2 \leq \sum\limits_{j=1}^{u} \frac{\alpha_{r_j}}{\alpha_0}
\end{equation}
\noindent in which $\rho_2$ is a uniform random variable with values in the range $(0,1)$.

At each time step ($t$), a time lapse ($\tau_t$) is derived based on $\alpha_0$, and is represented as:
\begin{equation}
\alpha_0 \cdot \tau_t = \mbox{ln} \frac{1}{\rho_1}
\end{equation}
\noindent in which $\rho_1$ is a uniform random variable with values in the range $(0,1)$. This process ends when $\sum\limits_{i=0}^{|T|} \tau_i < t_\theta$, where $T$ is the set of $t$ and $t_\theta$ is the maximum simulation time.




\begin{table}
	\centering
	\caption{Simulation parameters for astrocytes \cite{Lavrentovich:2008:jtheobio}.}
	\label{tb:paramvaluesast}
	\begin{tabular}{ c c  }
	    \toprule
	    Variable & Value  \\	    \midrule
	    $C_a$ & 0.1$\mu M$ \\
	    $E_a$ & 1.5$\mu M$ \\
	    $I_a$ & 0.1$\mu M$ \\
	    $\sigma_{0}$ & 0.05 $\mu M$  \\
	    $\kappa_{o}$ & 0.5 s$^{-1}$ \\
	    $\kappa_f$ &  0.5 s$^{-1}$\\
	    $\kappa_{d}$ &  0.08 s$^{-1}$\\
	    $\Sigma_{M2}$ &  15$\mu M/s$\\
	    $\kappa_2$ &  0.1$\mu M$\\
	    $\Sigma_p$ &  0.05$\mu M/s$\\
	    $\kappa_p$ &  0.3$\mu M$\\
	    $n$ &  2.02\\
	    $\kappa_{C1}$ & 0.15$\mu M$\\
	    $\kappa_{C2}$ &  0.15$\mu M$\\
	    $\kappa_I$ & 0.1$\mu M$\\
	    $\Sigma_{M3}$ & 40.0s$^{-1}$\\
	    $m$ & 2.2\\
	    $D$ & 350$\mu m^2/s$\\
	    \bottomrule
	  \end{tabular}
\end{table}

\section*{Logic gate model}

The synthetic logic gates programming is made on the reaction-diffusion process that governs the Ca$^{2+}$ signalling-based molecular communications model. We basically analyse the molecular concentration of the cell in order to defined the Ca$^{2+}$ concentration threshold that triggers the output of the logic gate. This approach is inspired by \cite{Stetter:2006:bionetics,Hiratsuka:1999:ieeetrans}, where a recurrent biophysical signalling pathway model was used for the logic gate design. 
The stages for logic gate operation are as follows: First, the Ca$^{2+}$ concentration threshold is defined for a specific logic operation function followed by the synthetic gene implementation. Secondly, the upcoming Ca$^{2+}$ from neighbouring cells through intercellular signalling are considered as inputs. In our case, we have two inputs from two different cells. All inputs are being transmitted to the logic gate cell during the \textit{signalling period} ($T_b$). The inputs interfere with the concentration in the cell cytosol $[C_a]$ alongside with the existing intracellular Ca$^{2+}$ signalling. Logic gates have two known states, an "OFF" state or "0" and an "ON" state or "1". The transition between states is performed when $[C_a]$ $>$ Ca$^{2+}$ concentration threshold for the "ON" state and $[C_a]$ $\leq$ Ca$^{2+}$ concentration threshold for the "OFF" state.

\section*{Static Timing Analysis}

Our analysis on the delay is based on conventional digital circuit static timing analysis. Based on a value of delay that refers to the time a logic gate gives a complete output, $d$, and a number of operations that is established, $n$, our total delay of operation time $D$ is the linear progression of the given initial delay value, which equals to

\begin{equation}
    D = \sum_{0}^{n} d = d * n
\end{equation}





\bibliography{References}

\end{document}